\documentclass[reprint,amsmath,amssymb,aps,pra]{revtex4-2}
\usepackage{graphicx}
\usepackage{booktabs}

\usepackage{dcolumn}
\usepackage{enumitem} 

\usepackage{bm}
\usepackage{mathrsfs}
\usepackage{amsmath}
\usepackage{mathtools}
\usepackage{braket}
\usepackage{siunitx}

\DeclareSIUnit{\atomicunit}{a.u.}
\DeclareSIUnit{\electronvolt}{eV}

\usepackage{physics}

\usepackage[hidelinks]{hyperref}

\usepackage[]{xcolor}
\usepackage{soul}

\begin{document}

\title{Non-equilibrium cavity pumping of electronic molecular polaritons}

\author{Yassir El Moutaoukal}
\author{Rosario R. Riso}
\author{Henrik Koch}
\email{henrik.koch@ntnu.no}

\affiliation{\vspace{4mm}Department of Chemistry, Norwegian University of Science and Technology, 7491 Trondheim, Norway}

\begin{abstract}
Strong light-matter coupling in optical cavities offers a non-invasive route to modify molecular properties, and it is usually reached by collectively coupling many molecules to the same mode.
Here we consider the photonic counterpart of the collective regime, in which a single molecule interacts with a cavity mode holding many photons, a steady state of the pumped cavity.
Since the eigenstates of the Pauli-Fierz Hamiltonian carry no transverse electric field, such a state cannot be described as an excited state of the light-matter system.
We therefore constrain the photon number of the strong coupling QED Hartree-Fock wave function with a Lagrange multiplier and self consistently minimize the energy with respect to all parameters.
The constrained reference carries the field of a coherent state with an average number of photons, generated by a cavity mode whose effective free-field frequency is lowered by the multiplier.
We apply this framework to benzene and a benzene-water complex, where the pumped cavity polarizes the molecules and modifies their interaction as with a static classical field.
For hydrogen peroxide, the field can reshape the torsional potential and the far-infrared torsional spectrum.
The classical driven field limit thus emerges from a quantized mean field description, and the photon density becomes a control parameter for cavity-modified chemistry.
\end{abstract}

\keywords{cavity quantum electrodynamics; polaritonic chemistry; strong coupling; photon pumping}

\maketitle

\section{Introduction}\label{sec:introduction}
Placing molecules inside an optical cavity changes the electromagnetic environment they interact with.
When a molecular excitation and a confined mode of the field exchange energy faster than either of them loses it, light and matter hybridize into polaritons~\cite{ebbesen2016hybrid,dovzhenko2018light,herrera2020molecular,garcia2021manipulating}.
These hybrid states inherit features from both constituents and their properties can be tuned through the frequency, the volume and the polarization of the optical device~\cite{latini2019cavity,latini2021ferroelectric}.
Experiments have shown that within the strong coupling regime photochemical reactions~\cite{Ebbesen_1,lee2024controlling,sasaki2025optical} and ground state reactivity~\cite{thomas2016ground,thomas2019tilting,ahn2023modification} can be modified, and \textit{ab initio} calculations predict further effects such as cavity control of selectivity~\cite{vu2022enhanced}.
These findings have given rise to polaritonic chemistry~\cite{feist2018polaritonic,ruggenthaler2023understanding,mandal2023theoretical}, whose promise is to use the cavity as a tunable and non-invasive handle on chemical processes.

The simplest description of strong coupling is the Jaynes-Cummings model~\cite{jaynes1963comparison,shore1993jaynes}.
This model retains a single cavity mode and reduces the molecule to a two-level system, a ground state $|g\rangle$ and an excited state $|e\rangle$ separated in energy by $\omega_{eg}$,
\begin{equation}\label{eq:jaynes_cummings}
H_{\mathrm{JC}} = \omega_{eg}|e\rangle\langle e| + \omega b^\dagger b + g\big(|e\rangle\langle g|\,b + |g\rangle\langle e|\,b^\dagger\big) ,
\end{equation}
\begin{figure}[b]
\centering
\includegraphics[width=\columnwidth]{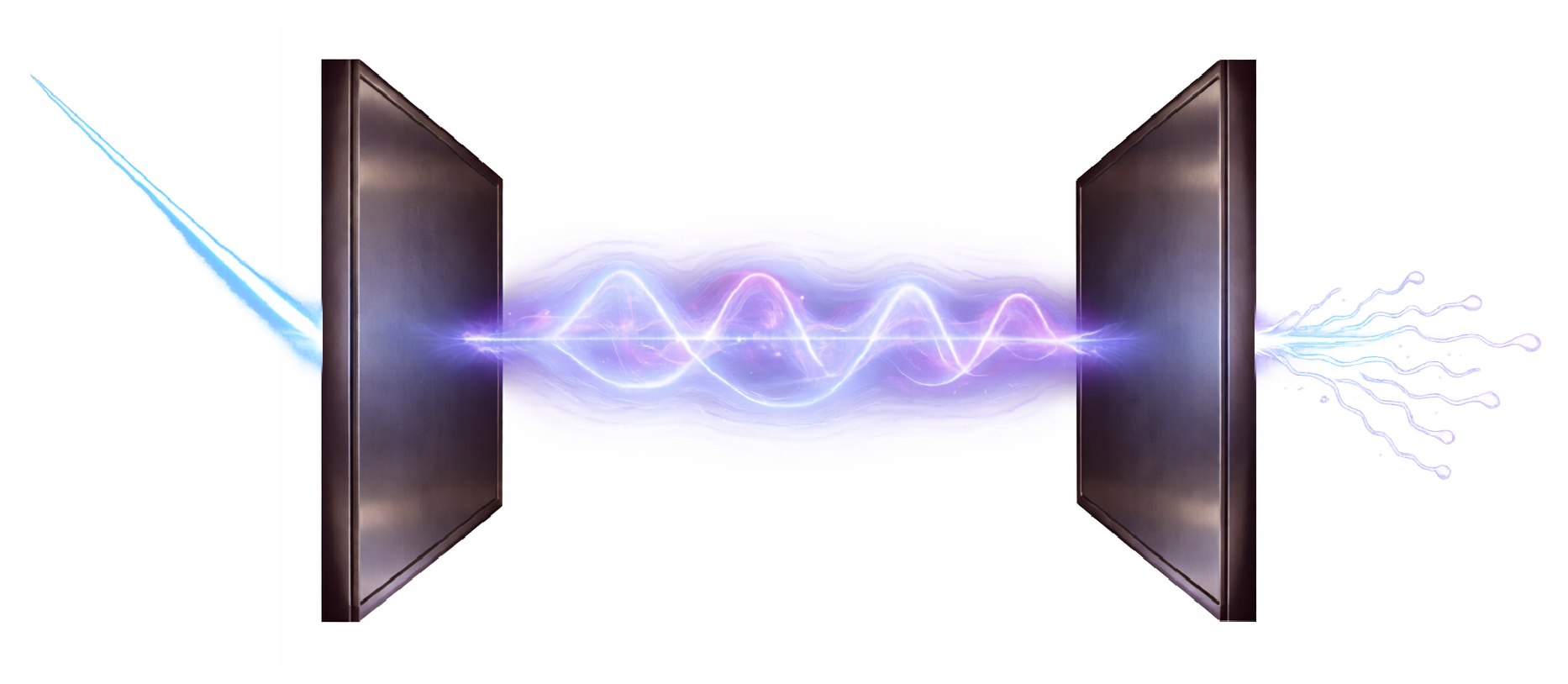}
\caption{Schematic representation of a pumped cavity.
An external source injects photons into the cavity (left), and photons leak out through the mirrors (right).
When injection and leakage balance, the cavity mode reaches a steady state with a fixed mean number of photons.}
\label{fig:pumped_cavity}
\end{figure}
where $b^\dagger$ and $b$ create and annihilate a photon of frequency $\omega$.
The coupling $g=\lambda\sqrt{\omega/2}\,(\mathbf{d}\cdot\pmb{\epsilon})_{eg}$ is the product of the molecular transition dipole along the field polarization $\pmb{\epsilon}$ and the vacuum field of the mode, whose strength grows as the quantization volume $V$ shrinks.
Only terms that conserve the number of excitations are retained, so the excited molecule in the empty cavity, $|e\rangle|0\rangle$, couples only to the ground state molecule with one photon, $|g\rangle|1\rangle$.
These two states mix into a lower and an upper polariton, whose energies are separated by the vacuum Rabi splitting, which depends on the detuning $\Delta=\omega_{eg}-\omega$ between the bare light and matter excitations as
\begin{equation}\label{eq:rabi_vacuum}
\Omega_R = \sqrt{\Delta^2 + 4g^2} .
\end{equation}
Notice that this splitting is nonzero even though the cavity contains no photons, since it is produced by the quantum vacuum fluctuations of the electromagnetic field.
Strong coupling is reached when $\Omega_R$ exceeds the linewidths of the molecule and of the cavity~\cite{dovzhenko2018light}.
For a single molecule in a conventional cavity, $g$ is far smaller than these linewidths and single molecule strong coupling has been achieved only in plasmonic nanocavities that confine the field to nanometric volumes~\cite{chikkaraddy2016single,ojambati2019quantum}.
In most experiments strong coupling is instead reached collectively~\cite{ying2025collective}.
When $N_{\mathrm{mol}}$ of identical molecules couple to the same mode, as in the Tavis-Cummings model~\cite{tavis1968exact}, a single excitation is shared by the symmetric, also referred to as bright, combination of the molecular excitations, which couples to the mode with the enhanced strength $g\sqrt{N_{\mathrm{mol}}}$.
In the collective regime the Rabi splitting reads
\begin{equation}\label{eq:rabi_collective}
\Omega_R = \sqrt{\Delta^2 + 4g^2N_{\mathrm{mol}}}
\end{equation}
and grows with the square root of the number of molecules, that is, their concentration.
What cannot be reached by shrinking the cavity around one molecule is reached by filling it with many of them.
However, the coupling per molecule remains $g$ and the remaining $N_{\mathrm{mol}}-1$ dark states do not mix with the field.
For this reason, how the collective effect is shared by many molecules and modifies the individual molecules is still an open question in the community~\cite{martinez2018can,sidler2020polaritonic,castagnola2024collective,perez2024collective,castagnola2025realistic,krupp2026first}.
The same square root enhancement can be achieved within the Jaynes-Cummings model when the cavity holds photons.
The states $|e\rangle|N_{\mathrm{ph}}-1\rangle$ and $|g\rangle|N_{\mathrm{ph}}\rangle$, which share $N_{\mathrm{ph}}$ excitations, are coupled with strength $g\sqrt{N_{\mathrm{ph}}}$ and split by
\begin{figure}[t]
\centering
\includegraphics[width=\columnwidth]{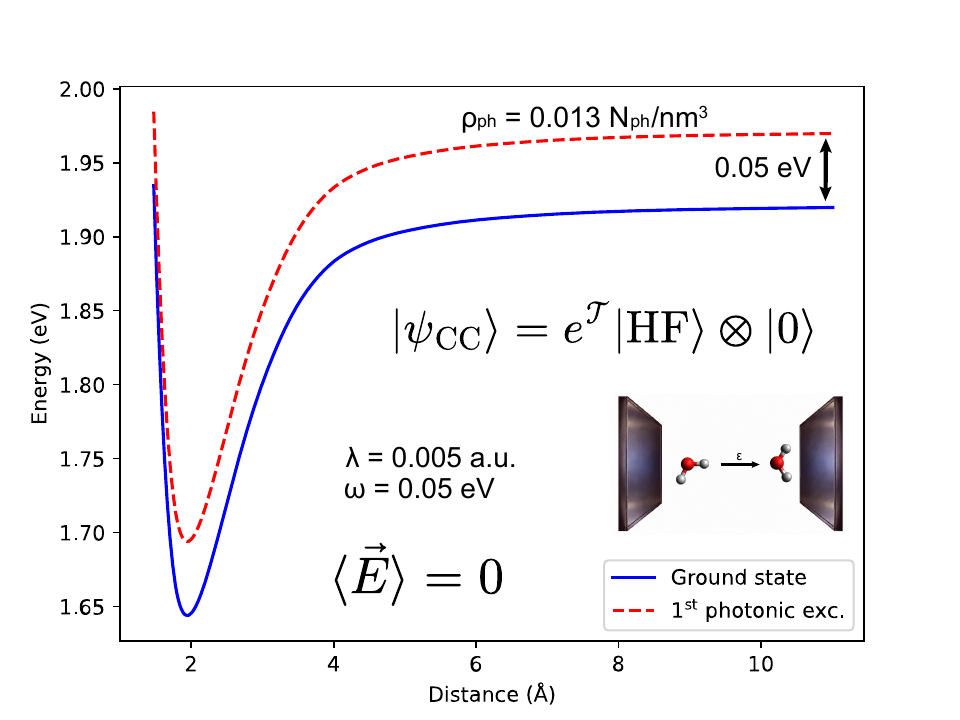}
\caption{Ground state (blue solid) and first photonic excitation (red dashed) of two hydrogen bonded water molecules as a function of their distance, from equation-of-motion QED-CCSD with $\lambda=0.005$~a.u., $\omega=0.05$~eV, and using an aug-cc-pVDZ basis set.
Both are eigenstates of Eq.~\ref{eq:pauli_fierz} and carry no mean electric field, $\langle\mathbf{E}\rangle=0$.}
\label{fig:qed_ccsd}
\end{figure}
\begin{equation}\label{eq:rabi_n}
\Omega_R = \sqrt{\Delta^2 + 4g^2 N_{\mathrm{ph}}} ,
\end{equation}
so that the photon number $N_{\mathrm{ph}}$ plays the role that the number of molecules plays in the Tavis-Cummings model.
This $\sqrt{N_{\mathrm{ph}}}$ scaling has been observed with Rydberg atoms and superconducting circuits~\cite{brune1996quantum,fink2008climbing}.
More specifically, photons can be accumulated in a cavity by pumping it with an external source, as sketched in Figure~\ref{fig:pumped_cavity}.
The pumped photons enter the cavity through the mirrors and leak out through them at the cavity loss rate.
Under continuous pumping, injection and leakage eventually balance and the mode reaches a steady state holding a fixed mean number of photons $N_{\mathrm{ph}}$, which grows with the square of the pump amplitude and decreases with the square of the loss rate~\cite{gardiner1985input,walls2008quantum}.
For a coherent pump, such as a laser, this steady state is a coherent state of the mode~\cite{glauber1963coherent,walls2008quantum}: the photon number fluctuates around $N_{\mathrm{ph}}$ and the field oscillates in time with a definite amplitude and phase.
A pumped cavity is thus none other than the photonic counterpart of the polaritonic collective regime: instead of many molecules sharing one photon, a single molecule interacts with many photons and the coupling is enhanced without shrinking the cavity.
The same enhancement is exploited in cavity optomechanics, where driving the cavity brings its coupling to a mechanical oscillator into the strong coupling regime~\cite{aspelmeyer2014cavity}, and it underlies the optomechanical description of Raman scattering from molecules in plasmonic cavities~\cite{roelli2016molecular}.
At large photon numbers, the splitting $2g\sqrt{N_{\mathrm{ph}}}$ coincides with the Rabi frequency $(\mathbf{d}\cdot\pmb{\epsilon})_{eg}E_0$ of a classical field with amplitude $E_0=\lambda\sqrt{2\omega N_{\mathrm{ph}}}$, and the enhancement provided by the photons approaches that of a classical driving field~\cite{sentef2020quantum,eckhardt2022quantum}.
The relation between the quantized and the classical descriptions of the cavity field has been examined with mixed quantum-classical treatments of the photons~\cite{hoffmann2019capturing,hoffmann2019benchmarking,rosenzweig2022analysis} and through the spectra of molecular polaritons~\cite{schwennicke2025molecular,simko2025twin}, and the response of molecular polaritons to external driving is attracting increasing attention~\cite{bustamante2026collective}.
The modifications in the electronic structure of molecules brought about by quantized fields cannot be rationalized within a two-level description and require a more detailed level of theory.

Describing how the molecular ground state $|g\rangle|0\rangle$ and its energy landscape are reshaped by the electromagnetic field~\cite{galego2015cavity,flick2017cavity,lacombe2019exact} requires at least an \textit{ab initio} optimization of the orbitals, and hence of the electron density, which determines the chemistry.
Capturing these effects requires the full electronic structure of the molecule to be coupled with the quantized field, as described by the Pauli-Fierz Hamiltonian~\cite{Photon+Atoms,craig1998molecular}.
Recently developed \textit{ab initio} cavity quantum electrodynamics (QED) methods~\cite{foley2023ab,weight2025ab} include density functional theory~\cite{tokatly2013time,ruggenthaler2014quantum,flick2017atoms}, coupled cluster~\cite{haugland2020coupled,mordovina2020polaritonic,fischer2026coherent}, perturbation theory~\cite{bauer2023perturbation,cui2024variational,el2025strong}, multiconfigurational and selected configuration interaction approaches~\cite{weight2023investigating,alessandro2025complete,zhang2026cqed}, and many-body Green's function methods~\cite{willow2026gw}.
At the mean field level, QED Hartree-Fock dresses a Slater determinant with a single coherent state of the photon field~\cite{haugland2020coupled}.
The strong coupling QED Hartree-Fock (SC-QED-HF) goes further and includes the coherent state dependence on the occupation of the molecular orbitals~\cite{riso2022molecular}, which entangles electrons and photons and captures the dependence of the molecular properties on the cavity frequency.
This orbital-specific approach has been extended to large molecules with improved convergence~\cite{el2024toward}, perturbative electron-photon correlation~\cite{el2025strong}, polaritonic response theory~\cite{castagnola2025strong}, and chiral cavities~\cite{el2026unveiling}.
All these methods target the eigenstates of the Pauli-Fierz Hamiltonian: the ground state of the molecule in the cavity vacuum and its polaritonic excited states.
A cavity holding photons might then be described by an excited state obtained through the photonic excitations within the QED equation of motion coupled cluster~\cite{haugland2020coupled} framework.
There, the excited states follow from the eigenvalues of the similarity transformed Hamiltonian and the cluster operator is composed of electronic, photonic, and coupled electronic-photonic excitations.
Figure~\ref{fig:qed_ccsd} shows the ground state and the first off-resonance photonic excitation for two hydrogen bonded water molecules.
Although the excited state contains a photon, it carries no electric field.
This holds for every eigenstate of the Pauli-Fierz Hamiltonian: in a stationary state the displacement field generated by the photons exactly balances the polarization of the molecule, and the transverse electric field vanishes exactly~\cite{mandal2020polarized}.
Like a Fock state, an eigenstate with photons has a well defined energy but no field associated with it.
On the other hand, the steady state of a pumped cavity is of a different kind.
It is maintained out of equilibrium by the pump and is not an eigenstate of the Hamiltonian, thus it carries a field that actively acts on the molecule beyond polarization effects.
Describing a molecule in a pumped cavity therefore requires targeting a non-equilibrium state with a prescribed average number of photons and a nonzero field, rather than an excited state.

In this work, we calculate such a steady-state at the mean field level.
Specifically, we constrain the average photon number of the SC-QED-HF wave function to $N_{\mathrm{ph}}$ with a Lagrange multiplier, and optimize the orbitals, the coherent state parameters and the multiplier self consistently.
In the cavity, the field of the pumped state oscillates in time with the phase of the pump.
We aim to model a snapshot of the state at the maximum of this oscillation, where the field is largest.
When the cavity frequency lies well below the electronic excitation energies, as in the systems studied here, the electrons follow the oscillating field adiabatically, and the snapshot describes their instantaneous state.
Working within the Born-Oppenheimer approximation, the nuclei move in the average field of the electrons.
We find that the pumped reference carries the electric field $\lambda\sqrt{2\omega N_{\mathrm{ph}}}$ of a coherent state with $N_{\mathrm{ph}}$ photons, and that the molecule responds to it as to a static classical field of the same strength.
The classical driven field limit thus emerges from the quantized description.
The photon density becomes the control parameter: cavities with the same $\lambda^2N_{\mathrm{ph}}$ give the same chemistry, so a large cavity holding many photons acts on a molecule as a small cavity holding few.
At first sight this is genuinely an effective light-matter coupling enhancement within the ground state, using photons out of resonance to any electronic excitation.
We illustrate these results for the polarization of a benzene molecule, the interaction between benzene and water, and finally the torsional normal mode potential and infrared spectrum of hydrogen peroxide.
Section~\ref{sec:theory} presents the theory, Section~\ref{sec:implementation} the implementation, Section~\ref{sec:results} the results, and in Section~\ref{sec:conclusions} we give our conclusions.

\section{Theory}\label{sec:theory}
Throughout this paper we work in atomic units and consider a molecule in a cavity supporting a single quantized mode of frequency $\omega$ and polarization $\pmb{\epsilon}$.
In the length gauge and within the dipole approximation, the light-matter system is described by the Pauli-Fierz Hamiltonian~\cite{Photon+Atoms,craig1998molecular}
\begin{equation}\label{eq:pauli_fierz}
H = H_e + \omega b^\dagger b - \lambda\sqrt{\frac{\omega}{2}}(\mathbf{d}\cdot\pmb{\epsilon})(b+b^\dagger) + \frac{\lambda^2}{2}(\mathbf{d}\cdot\pmb{\epsilon})^2 ,
\end{equation}
where $H_e$ is the electronic Hamiltonian, $\mathbf{d}$ is the molecular dipole operator, and $b^\dagger$ and $b$ respectively creates and annihilates a photon in the mode.
The light-matter coupling strength
\begin{equation}\label{eq:coupling}
\lambda = \sqrt{\frac{4\pi}{V}}
\end{equation}
is fixed by the quantization volume $V$ of the cavity.
The third term of Eq.~\ref{eq:pauli_fierz} is the bilinear interaction and correlates electrons and photons, while the last term is the dipole self energy (DSE), without which the Hamiltonian is unbounded from below~\cite{rokaj2018light,schafer2020relevance,hoffmann2020effect}.
The photonic terms can be collected into a single displaced oscillator
\begin{equation}\label{eq:pauli_fierz_square}
H = H_e + \omega\Big(b^\dagger - \frac{\lambda}{\sqrt{2\omega}}\mathbf{d}\cdot\pmb{\epsilon}\Big)\Big(b - \frac{\lambda}{\sqrt{2\omega}}\mathbf{d}\cdot\pmb{\epsilon}\Big) 
\end{equation}
which shows that the field is displaced proportionally to the molecular dipole and that this displacement is itself an electronic operator.
In the second quantization formalism~\cite{helgaker2014molecular}, the electronic Hamiltonian is
\begin{equation}
    H_e=\sum_{pq}h_{pq}E_{pq}+\frac{1}{2}\sum_{pqrs}g_{pqrs}e_{pqrs},
\end{equation}
written in terms of one and two-electron integrals ($h_{pq}$ and $g_{pqrs}$) and the singlet one and two-electron operators
\begin{equation}
\begin{split}
&E_{pq}=\sum_{\sigma}a^{\dagger}_{p\sigma}a_{q\sigma} \\ 
&e_{pqrs}=E_{pq}E_{rs}-\delta_{rq}E_{ps},
\end{split}
\end{equation}
where $a^{\dagger}_{p\sigma}$ and $a_{p\sigma}$ are the creation and annihilation operators for an electron in orbital $p$ and spin $\sigma$.
The interaction operator reads
\begin{equation}
    \mathbf{d}\cdot\pmb{\epsilon}=\sum_{pq}(\mathbf{d}\cdot\pmb{\epsilon})_{pq}E_{pq}.
\end{equation}
Throughout we work in the dipole basis that diagonalizes the interaction integrals
\begin{equation}\label{eq:dipole_basis}
\sum_{rs}C_{rp}(\mathbf{d}\cdot\pmb{\epsilon})_{rs}C_{sq} = (\tilde{\mathbf{d}}\cdot\pmb{\epsilon})_{pp}\delta_{pq} 
\end{equation}
and denote this by a tilde ($\sim$) on the relevant quantities.
The SC-QED-HF wave function is described by a reference state composed of a Slater determinant for the electrons and the photonic vacuum. The two are then entangled by an orbital-specific coherent state transformation, such that
\begin{equation}\label{eq:sc_wf}
|\psi_{\mathrm{SC}}\rangle = U_{\mathrm{SC}}\,\exp({\kappa}) \ |\mathrm{HF}\rangle\otimes|0\rangle ,
\end{equation}
where
\begin{equation}\label{eq:U_SC}
U_{\mathrm{SC}} = \exp\Bigg(-\frac{\lambda}{\sqrt{2\omega}}\sum_p \eta_p \tilde{E}_{pp}\big(b-b^\dagger\big)\Bigg) .
\end{equation}
In Eq.~\ref{eq:sc_wf} the non-redundant orbital rotations are collected in
\begin{equation}
    \kappa=\sum_{ai}\kappa_{ai}(E_{ai}-E_{ia})
\end{equation}
and the wave function is then composed of two sets of parameters: $\{\kappa_{ai}\}$ and $\{\eta_p\}$.
The ground state energy is the expectation value of the Pauli-Fierz Hamiltonian with the SC wave function
\begin{equation}\label{eq:E_SC}
\begin{split}
E_{\mathrm{SC}} &=  \sum_{pq}\tilde{h}_{pq}Q_{pq}\tilde{D}_{pq} + \frac{1}{2}\sum_{pqrs}\tilde{g}_{pqrs}Q_{pqrs}\tilde{d}_{pqrs} \\
& + \frac{\lambda^2}{2}\sum_{pq}(\tilde{\mathbf{d}}\cdot\pmb{\epsilon})^\eta_{pp}\,S_{pq}\,(\tilde{\mathbf{d}}\cdot\pmb{\epsilon})^\eta_{qq} ,
\end{split}
\end{equation}
where $\tilde{D}_{pq}$ and $\tilde{d}_{pqrs}$ are the one- and two-electron densities of the reference determinant. The two-electron integrals are scaled with the Gaussian factors defined as
\begin{equation}\label{eq:gaussian factors}
\begin{split}
& Q_{pqrs} = \exp\Big(-\frac{\lambda^2}{4\omega}\Delta_{pqrs}^2\Big) \\
& \Delta_{pqrs} = \eta_p - \eta_q + \eta_r - \eta_s
\end{split}
\end{equation}
and same for the one-electron integrals with $Q_{pq}$ and $\Delta_{pq}$ defined analogously.
The Gaussian factors damp an integral whenever the orbitals it connects carry different coherent state parameters, and they are responsible for the frequency dependence at the mean field level~\cite{el2025strong}.
The last term of Eq.~\ref{eq:E_SC} is the self energy, where the interaction integrals are shifted by the $\eta$-parameters
\begin{equation}\label{eq:shifted_dipole}
    (\tilde{\mathbf{d}}\cdot\pmb{\epsilon})^\eta_{pp} = (\tilde{\mathbf{d}}\cdot\pmb{\epsilon})_{pp} - \eta_p .
\end{equation}
These integrals enter the energy equation in a quadratic form with the Gram matrix of the operators $\tilde{E}_{pp}$ acting on the Hartree Fock reference
\begin{equation}\label{eq:S}
S_{pq} = \langle\tilde{E}_{pp}\tilde{E}_{qq}\rangle = \tilde{D}_{pp}\delta_{pq} + \tilde{D}_{pp}\tilde{D}_{qq} - \frac{1}{2}\tilde{D}_{pq}\tilde{D}_{qp} .
\end{equation}
The matrix $S$ is symmetric, positive semidefinite and it reappears in the photon number equation
\begin{equation}\label{eq:photon_number}
\langle b^\dagger b\rangle_{\mathrm{SC}} = \frac{\lambda^2}{2\omega}\sum_{pq}\eta_p S_{pq}\eta_q ,
\end{equation}
which turns out to be a quadratic form in the coherent state parameters.
For a common $\eta_p = \eta$ the photon number equation reduces to $\lambda^2\eta^2N_e^2/2\omega$, with $N_e$ the number of electrons.
The photons in the reference are thus entirely determined by the displacement, which is in turn tied to the electronic structure.

The mean electromagnetic fields follow from the same expectation values.
In atomic units the transverse electric field is
\begin{equation}\label{electric field}
    \mathbf{E}=4\pi(\mathbf{D}-\mathbf{P}),
\end{equation}
where the displacement and polarization fields are~\cite{craig1998molecular,Photon+Atoms}
\begin{equation}\label{eq:D_field}
\begin{split}
& \mathbf{D} = \frac{\lambda\sqrt{\omega}}{4\pi\sqrt{2}}\langle b+b^\dagger\rangle\,\pmb{\epsilon}, \\
& \mathbf{P} = \frac{\lambda^2}{4\pi}\langle\mathbf{d}\cdot\pmb{\epsilon}\rangle\,\pmb{\epsilon} .
\end{split}
\end{equation}
For any exact eigenstate of Eq.~\ref{eq:pauli_fierz} the expectation value of the commutator $[b,H]=\omega b-\lambda\sqrt{\omega/2}\,\mathbf{d}\cdot\pmb{\epsilon}$ vanishes, because $H$ acting on the eigenstate from either left or right returns the same real energy.
This gives $\langle b\rangle=\lambda\langle\mathbf{d}\cdot\pmb{\epsilon}\rangle/\sqrt{2\omega}$ and therefore $\mathbf{D}=\mathbf{P}$. Thus a stationary state carries no transverse field, regardless of how many photons it contains.
The SC-QED-HF reference is not an exact eigenstate and its field has to be evaluated explicitly.
The transformation in Eq.~\ref{eq:U_SC} displaces the bosonic operators such that $\langle b\rangle=\lambda\langle X\rangle/\sqrt{2\omega}$ with
\begin{equation}\label{X_avg}
    \langle X\rangle=\sum_p\eta_p\tilde{D}_{pp}.
\end{equation}
Considering that
\begin{equation}\label{expect dipole}
\langle\mathbf{d}\cdot\pmb{\epsilon}\rangle=\sum_p(\tilde{\mathbf{d}}\cdot\pmb{\epsilon})_{pp}\tilde{D}_{pp}
\end{equation}
and inserting both Eqs.~\ref{X_avg} and \ref{expect dipole} in Eq.~\ref{electric field} gives
\begin{equation}\label{eq:field}
\mathbf{E} = -\lambda^2\sum_p(\tilde{\mathbf{d}}\cdot\pmb{\epsilon})^\eta_{pp}\tilde{D}_{pp}\,\pmb{\epsilon} .
\end{equation}
So, the electric field is proportional to the shifted dipole of Eq.~\ref{eq:shifted_dipole} averaged over the reference.
A cavity that is pumped holds a prescribed number of photons and we impose this with a constraint.
We define the Lagrangian
\begin{equation}\label{eq:lagrangian}
\mathcal{L} = \langle\psi_{\mathrm{SC}}|H|\psi_{\mathrm{SC}}\rangle + \mu\big(\langle\psi_{\mathrm{SC}}|b^\dagger b|\psi_{\mathrm{SC}}\rangle - N_{\mathrm{ph}}\big) ,
\end{equation}
where the multiplier $\mu$ enforces $\langle b^\dagger b\rangle_{\mathrm{SC}}=N_{\mathrm{ph}}$.
Adding $\mu b^\dagger b$ to Eq.~\ref{eq:pauli_fierz} replaces the photon frequency by $\omega+\mu$ and leaves the coupling untouched, so the constrained reference is the SC-QED-HF state of a mode of effective free-field frequency $\omega+\mu$, displaced by the molecule until it holds an average of $N_{\mathrm{ph}}$ photons.
The constraint fixes the mean photon number, while the state of the mode is set by the coherent state parametrization of Eq.~\ref{eq:sc_wf}: a Fock or a thermal state with the same photon number would have $\langle b\rangle=0$ and carry no field beyond the molecular polarization.
The stationary points of Eq.~\ref{eq:lagrangian} are found from its derivatives with respect to the two sets of parameters and the Lagrange multiplier.
Differentiating with respect to the multiplier returns the constraint,
\begin{equation}\label{eq:mu_gradient}
\frac{\partial\mathcal{L}}{\partial\mu} = \langle b^\dagger b\rangle_{\mathrm{SC}} - N_{\mathrm{ph}} .
\end{equation}
The coherent state $\eta$-parameters enter the Gaussian factors, the shifted dipole integrals and the photon number, and the gradient is
\begin{equation}\label{eq:eta_gradient}
\begin{split}
\frac{\partial\mathcal{L}}{\partial\eta_t} &= \sum_{pq}\tilde{h}_{pq}\tilde{D}_{pq}\frac{\partial Q_{pq}}{\partial\eta_t} + \frac{1}{2}\sum_{pqrs}\tilde{g}_{pqrs}\tilde{d}_{pqrs}\frac{\partial Q_{pqrs}}{\partial\eta_t} \\
& - \lambda^2\sum_q S_{tq}(\tilde{\mathbf{d}}\cdot\pmb{\epsilon})^\eta_{qq} + \frac{\lambda^2\mu}{\omega}\sum_q S_{tq}\eta_q ,
\end{split}
\end{equation}
where the derivatives of the Gaussian factors are
\begin{equation}\label{eq:Q_derivative}
\begin{split}
&\frac{\partial Q_{pqrs}}{\partial\eta_t} = -\frac{\lambda^2}{2\omega}\Delta_{pqrs}\,\sigma^t_{pqrs}\,Q_{pqrs} \\
& \sigma^t_{pqrs}=\delta_{pt}-\delta_{qt}+\delta_{rt}-\delta_{st}
\end{split}
\end{equation}
and correspondingly for $Q_{pq}$.
The first two terms of Eq.~\ref{eq:eta_gradient} makes the coherent state parameters more similar since the Gaussian factors are largest when they are equal, while the third term forces them towards the dipole integrals.
The last term is the only one that carries the constraint and it is proportional to $\mu$.
The orbital rotations enter through the density matrix, where stationarity is given by the generalized Brillouin condition $\tilde{F}_{ai}=0$. The Fock matrix of the Lagrangian is the standard SC-QED-HF Fock matrix~\cite{riso2022molecular,el2024toward} augmented by the derivative of the constraint,
\begin{equation}\label{eq:fock_mu}
\tilde{F}^{\,\mu}_{pq} = \frac{\lambda^2\mu}{2\omega}\Big[\big(\eta_p^2 + 2\eta_p\langle X\rangle\big)\delta_{pq} - \eta_p\eta_q\tilde{D}_{pq}\Big] ,
\end{equation}
where $\langle X\rangle$ is the mean displacement defined in Eq.~\ref{X_avg}.
The photonic $\eta$-parameters and the multiplier are strongly coupled and they are best treated together.
Collecting the second derivatives of Eq.~\ref{eq:lagrangian} with respect to $\eta$ and $\mu$ gives the augmented Hessian
\begin{equation}\label{eq:augmented_hessian}
\mathbf{E}^{(2)} =
\begin{pmatrix}
\mathbf{H}^{\eta\eta} & \mathbf{w} \\[2pt]
\mathbf{w}^{T} & 0
\end{pmatrix} ,
\end{equation}
where the diagonal block is
\begin{equation}\label{eq:hessian_block}
H^{\eta\eta}_{tu} = H^{\mathrm{el}}_{tu} + \lambda^2\Big(1+\frac{\mu}{\omega}\Big)S_{tu} ,
\end{equation}
and off-diagonal block given by the derivative of the photon number,
\begin{equation}\label{eq:border}
w_t = \frac{\partial^2\mathcal{L}}{\partial\eta_t\partial\mu} = \frac{\lambda^2}{\omega}\sum_q S_{tq}\eta_q .
\end{equation}
In Eq.~\ref{eq:hessian_block} the term $H^{\mathrm{el}}_{tu}$ collects the second derivatives of the Gaussian factors,
\begin{equation}\label{eq:Q_second_derivative}
\frac{\partial^2 Q_{pqrs}}{\partial\eta_t\partial\eta_u} = \frac{\lambda^2}{2\omega}\Big[\frac{\lambda^2}{2\omega}\Delta_{pqrs}^2-1\Big]\sigma^t_{pqrs}\,\sigma^u_{pqrs}\,Q_{pqrs} 
\end{equation}
and analogously for $Q_{pq}$, weighted by the corresponding integrals and densities.
The structure of Eq.~\ref{eq:augmented_hessian} has two consequences.
The lower diagonal element vanishes because the Lagrangian is linear in the multiplier $\mu$, such that the augmented Hessian is a saddle point problem and is indefinite by construction.
Moreover, the constraint multiplies the photonic curvature in Eq.~\ref{eq:hessian_block} by $(\omega+\mu)/\omega$, this flattens that curvature as $\mu$ approaches $-\omega$ and reverses its sign below, such that the pumped reference is a minimum only on the constraint surface.
Interestingly, the electric field carried by the constrained reference can be obtained in a closed form.
At a stationary point every gradient component of Eq.~\ref{eq:eta_gradient} vanishes and so does their sum over $t$.
In this sum the first line of Eq.~\ref{eq:eta_gradient} drops out because $\sum_t\sigma^t_{pqrs}=0$ and likewise for the one-electron factors.
In the second line, the columns of $S$ sum to
\begin{equation}
    \sum_pS_{pq}=N_e\tilde{D}_{qq} ,
\end{equation}
since $\sum_p\tilde{E}_{pp}$ is the number operator and the reference is an eigenstate. In Eq.~\ref{eq:S} the one-electron term cancels the last exchange term through the idempotency $\tilde{\mathbf{D}}^2=2\tilde{\mathbf{D}}$ of the closed shell density.
The summed stationarity condition then relates the shifted dipole to the displacement,
\begin{equation}\label{eq:uniform_shift}
\sum_p(\tilde{\mathbf{d}}\cdot\pmb{\epsilon})^\eta_{pp}\tilde{D}_{pp} = \frac{\mu}{\omega}\langle X\rangle .
\end{equation}
The left-hand side equals $\langle\mathbf{d}\cdot\pmb{\epsilon}\rangle-\langle X\rangle$, which gives $\langle X\rangle=\omega\langle\mathbf{d}\cdot\pmb{\epsilon}\rangle/(\omega+\mu)$, and inserting both results into Eq.~\ref{eq:field} yields
\begin{equation}\label{eq:field_mu}
\mathbf{E} = -\frac{\lambda^2\mu}{\omega+\mu}\langle\mathbf{d}\cdot\pmb{\epsilon}\rangle\,\pmb{\epsilon} .
\end{equation}
Without the constraint $\mu$ vanishes and so does the field, in agreement with the general result for eigenstates. Thus the unconstrained SC-QED-HF ground state does not carry an electric field.
A field only appears when the photon number is constrained, and it grows as $\mu$ approaches $-\omega$, that is, as the effective free-field frequency of the mode is driven to zero.
It is worth noticing that when the variance $\langle X^2\rangle=\sum_{pq}\eta_pS_{pq}\eta_q$ over the reference is negligible, 
\begin{equation}
    \sum_{pq}\eta_pS_{pq}\eta_q\simeq\langle X\rangle^2,
\end{equation}
the electric field reduces to
\begin{equation}\label{eq:field_classical}
\mathbf{E} \simeq \Big(\lambda\sqrt{2\omega N_{\mathrm{ph}}} - \lambda^2\langle\mathbf{d}\cdot\pmb{\epsilon}\rangle\Big)\pmb{\epsilon} .
\end{equation}
This field corresponds to a coherent state with $N_{\mathrm{ph}}$ photons~\cite{glauber1963coherent} and reduced by the polarization of the molecule, which scales as $\lambda^2$, that is negligible for small coupling strengths.
The photon number therefore acts as a dial for the field seen by the molecule and the way this field is generated is examined in Section~\ref{sec:conclusions}.
In Appendix~\ref{app:statistics} the photon statistics of the reference beyond the mean photon number show that the described light is classical by construction.
In Appendix~\ref{app:squeezing} we describe the extension of the parametrization to squeezed states, which is required to describe nonclassical light.

\section{Implementation}\label{sec:implementation}
The parameters of the wave function are obtained by self consistently solving the stationarity conditions of Section~\ref{sec:theory}.
The orbitals are updated by diagonalizing the Fock matrix, as in the unconstrained theory~\cite{riso2022molecular,el2024toward}, while the coherent state parameters and the multiplier are updated together with a Newton step,
\begin{equation}\label{eq:newton_step}
\begin{pmatrix}\Delta\pmb{\eta}\\ \Delta\mu\end{pmatrix} = -\begin{pmatrix}
\mathbf{H}^{\eta\eta} & \mathbf{w} \\[2pt]
\mathbf{w}^{T} & 0
\end{pmatrix}^{-1}\begin{pmatrix}\partial\mathcal{L}/\partial\pmb{\eta}\\ \partial\mathcal{L}/\partial\mu\end{pmatrix} ,
\end{equation}
with the augmented Hessian of Eq.~\ref{eq:augmented_hessian}.
The last row of Eq.~\ref{eq:newton_step} reads $\mathbf{w}^T\Delta\pmb{\eta}=N_{\mathrm{ph}}-\langle b^\dagger b\rangle_{\mathrm{SC}}$, and since $\mathbf{w}$ is the gradient of the photon number, each step brings the photon number back to $N_{\mathrm{ph}}$ to first order in $\Delta\pmb{\eta}$.
A displacement $\mathbf{v}$ for which $\sum_p v_p\tilde{E}_{pp}|\mathrm{HF}\rangle$ vanishes leaves the wave function unchanged, and because the norm of this vector is $\mathbf{v}^T\mathbf{S}\mathbf{v}$, such directions are exactly the null space of the matrix $\mathbf{S}$ of Eq.~\ref{eq:S}.
These redundant displacements make $\mathbf{H}^{\eta\eta}$ singular and we remove them by diagonalizing $\mathbf{S}$, treating eigenvalues below $10^{-7}$ as zero, and projecting the gradient and the Hessian onto the complement of the corresponding eigenvectors~\cite{el2026unveiling}.
The projected Hessian is then level shifted by $10^{-6} \ \mathrm{a.u.}$ to ensure invertibility.
The off-diagonal blocks require no projection as $\mathbf{w}$ is proportional to $\mathbf{S}\pmb{\eta}$ and $\mathbf{S}$ is symmetric, and $\mathbf{v}^T\mathbf{w}=0$ for every redundant direction $\mathbf{v}$, which means that moving along a redundant direction leaves the photon number unchanged.
The coherent state $\eta$-parameters are initialized to the dipole integrals $(\tilde{\mathbf{d}}\cdot\pmb{\epsilon})_{pp}$, their infinite coupling value, and are then all shifted by the same amount so that Eq.~\ref{eq:photon_number} equals $N_{\mathrm{ph}}$ already in the first iteration.
This condition is quadratic in the shift and among the two roots we retain the one that displaces the field along $+\pmb{\epsilon}$; reversing $\pmb{\epsilon}$ gives the opposite orientation of the field relative to the molecule.
Each iteration of the self consistent field procedure builds the Fock matrix in the dipole basis, updates the orbitals with a Roothaan-Hall step accelerated by the direct inversion in the iterative subspace (DIIS) algorithm~\cite{diis1,diis2}. The $\eta$ and $\mu$ are updated with one step of Eq.~\ref{eq:newton_step}.
The method is implemented in a development version of the e$^\mathcal{T}$ program~\cite{eT}.

\section{Results and discussion}\label{sec:results}
All calculations use a single cavity mode of frequency $\omega=1.36$~eV, well below the electronic excitation energies of the molecules studied, and coupling strengths $\lambda=0.001$ and $0.005$~a.u., which correspond through Eq.~\ref{eq:coupling} to quantization volumes of 1862~nm$^3$ and 74.5~nm$^3$.
For the intermolecular interaction calculations in Sections~\ref{sec:benzene_water} and~\ref{sec:photon_density} we used the cc-pVDZ basis set~\cite{dunning1989a}, while we adopted the aug-cc-pVDZ basis set~\cite{kendall1992electron} for the density difference analyses in Sections~\ref{sec:cda} and~\ref{sec:static} as well as for the intramolecular torsional potential of hydrogen peroxide in Section~\ref{sec:h2o2}.
All the static field calculations use the Hartree-Fock Hamiltonian
\begin{equation}\label{eq:static_field}
H = H_e - E_0\,\mathbf{d}\cdot\pmb{\epsilon}
\end{equation}
with the field strength $E_0$ of Eq.~\ref{eq:field_mu} obtained from the corresponding pumped calculation.
We refer to the unconstrained SC-QED-HF ground state, with $N_{\mathrm{ph}}\approx0$, as the dark cavity where only the virtual photons that accompany the molecular polarization are present.
As discussed in Section~\ref{sec:introduction}, the pumped reference describes the state of the system at the maximum of the field oscillation.
The electrons follow the field adiabatically, while for the nuclear properties in Sections~\ref{sec:benzene_water}--\ref{sec:h2o2} we interpret the pumped reference as a cavity with a static bias, according with the Born-Oppenheimer approximation.

\subsection{Induced polarization of benzene}\label{sec:cda}
\begin{figure*}[t]
\centering
\includegraphics[width=0.50\textwidth]{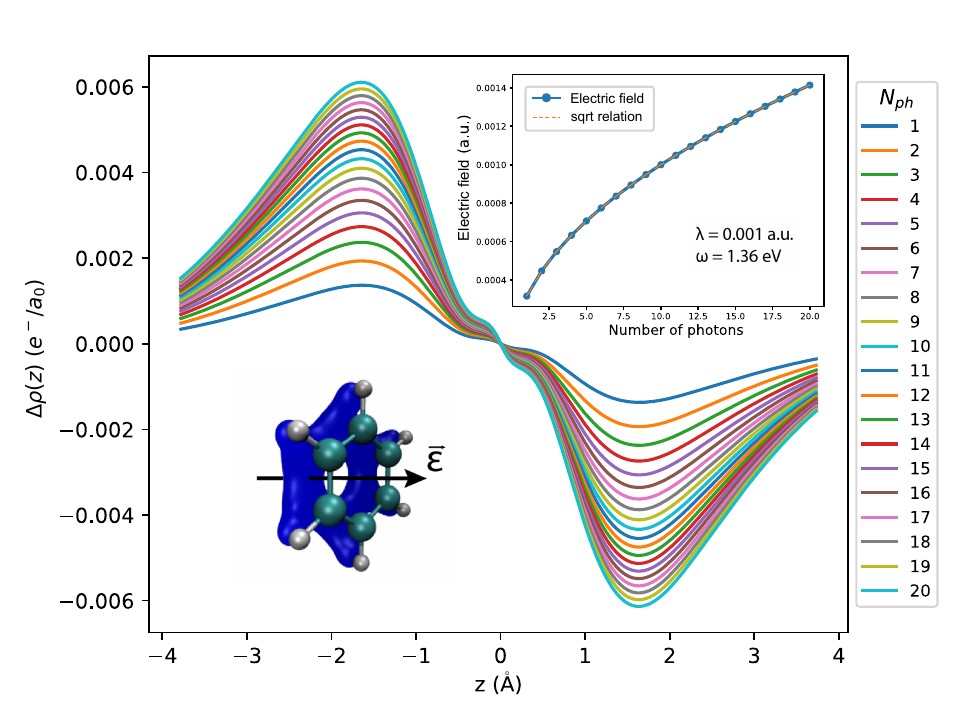}\hfill
\includegraphics[width=0.48\textwidth]{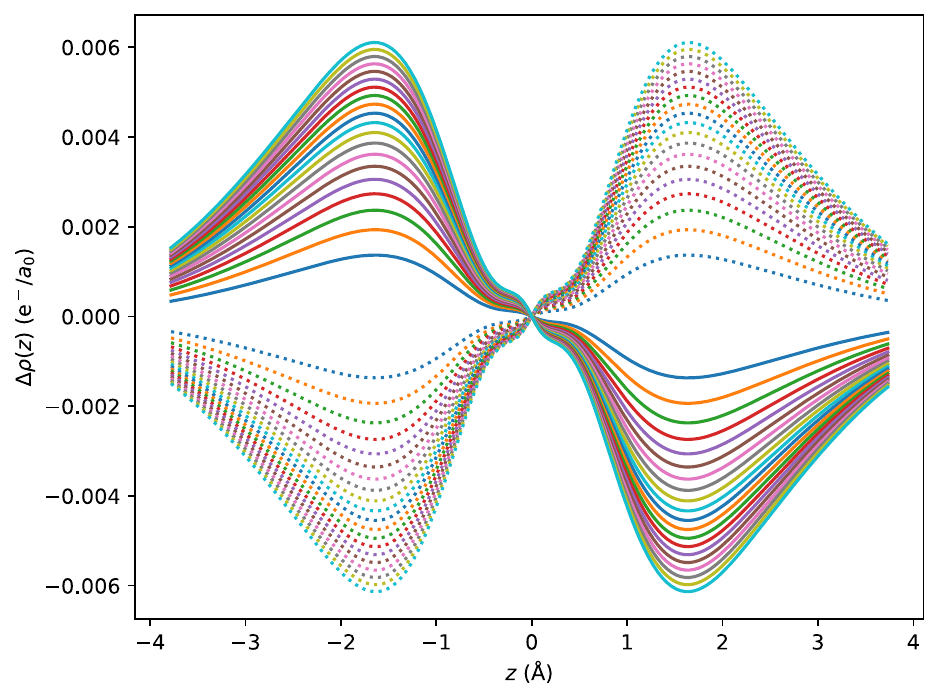}
\caption{Plane-integrated density difference $\Delta\rho(z)$ in Eq.~\ref{eq:planar_density} for benzene in a pumped cavity with $\lambda=0.001$~a.u.\ and $\omega=1.36$~eV, for $N_{\mathrm{ph}}$ from 1 to 20.
Left: field polarized along $+z$; the inset shows the electric field strength as a function of the photon number together with the relation $\lambda\sqrt{2\omega N_{\mathrm{ph}}}$.
For the centrosymmetric benzene molecule $\Delta\rho(z)$ is odd in $z$.
Right: field along $+z$ (solid) and $-z$ (dotted).
Reversing $\pmb{\epsilon}$ changes the sign of $\Delta\rho(z)$.}
\label{fig:cda}
\end{figure*}

We first consider how the pumped field polarizes a benzene molecule with the field polarized along the $z$ axis orthogonal to the molecular plane.
To visualize the charge rearrangement, we integrate the difference between the electron densities of the pumped reference and of the Hartree-Fock ground state over planes perpendicular to the field and parallel to the ring,
\begin{equation}\label{eq:planar_density}
\Delta\rho(z) = \int\!\!\int\big[\rho_{\mathrm{pump}}(x,y,z)-\rho_{\mathrm{HF}}(x,y,z)\big]\,dx\,dy ,
\end{equation}
which gives the change in electronic charge per unit length along $z$.
Figure~\ref{fig:cda} shows $\Delta\rho(z)$ for photon numbers between 1 and 20 at $\lambda=0.001$~a.u.
Electrons accumulate at negative $z$ and are depleted at positive $z$, since, being negatively charged, they move against the field so the pumped cavity induces a dipole along the polarization.
The amplitude of $\Delta\rho(z)$ grows as $\sqrt{N_{\mathrm{ph}}}$, following the field strength shown in the inset, which is given by Eq.~\ref{eq:field_classical} when the variance of $X$ is negligible, as expected for a density that responds linearly to the field.
The linear character of the response is also confirmed in the right panel, where reversing the polarization changes the sign of $\Delta\rho(z)$.

\subsection{Pumped cavity and static field}\label{sec:static}
\begin{figure}[h]
\centering
\includegraphics[width=\columnwidth]{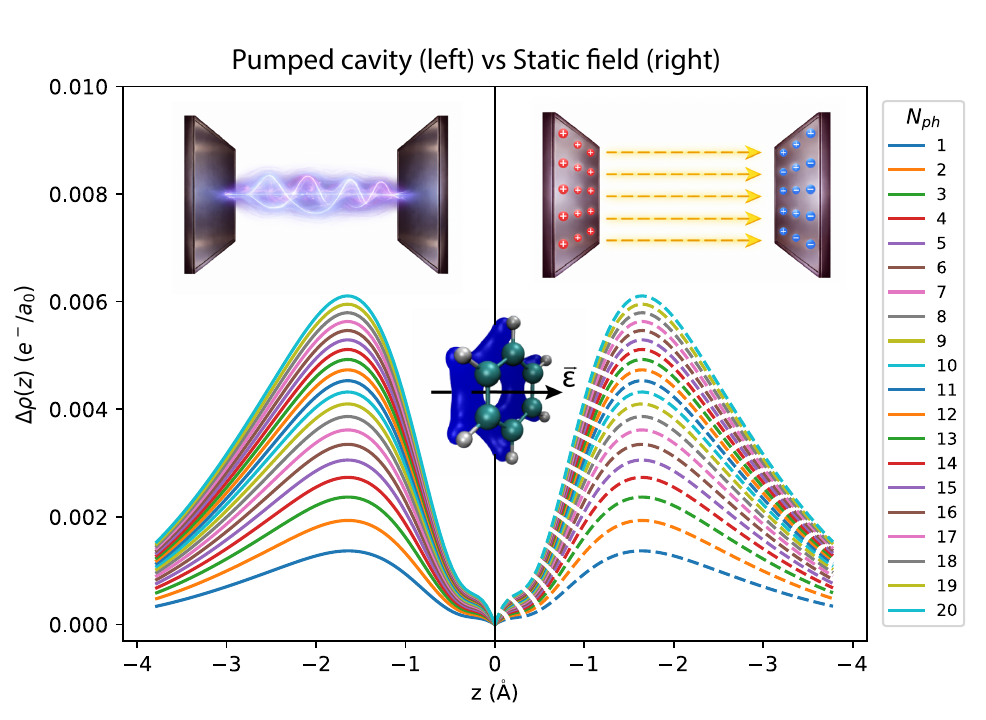}
\caption{Plane-integrated density difference $\Delta\rho(z)$ for benzene in the pumped cavity (left) and in a static field of the same strength (right), for $N_{\mathrm{ph}}$ from 1 to 20, with $\lambda=0.001$~a.u.\ and $\omega=1.36$~eV.}
\label{fig:static}
\end{figure}

Figure~\ref{fig:static} compares $\Delta\rho(z)$ in the pumped cavity with the density difference produced by a static field of the same strength, Eq.~\ref{eq:static_field}.
The two sets of curves coincide on the scale of the figure for every photon number.
The pumped SC-QED-HF reference therefore polarizes the molecule as a classical static field of strength $\lambda\sqrt{2\omega N_{\mathrm{ph}}}$.
This is the regime of Eq.~\ref{eq:field_classical} in which the variance of the displacement is negligible compared to its mean and the molecule experiences the mean field of the mode.

\subsection{Benzene-water interaction}\label{sec:benzene_water}
\begin{figure*}[t]
\centering
\includegraphics[width=0.49\textwidth]{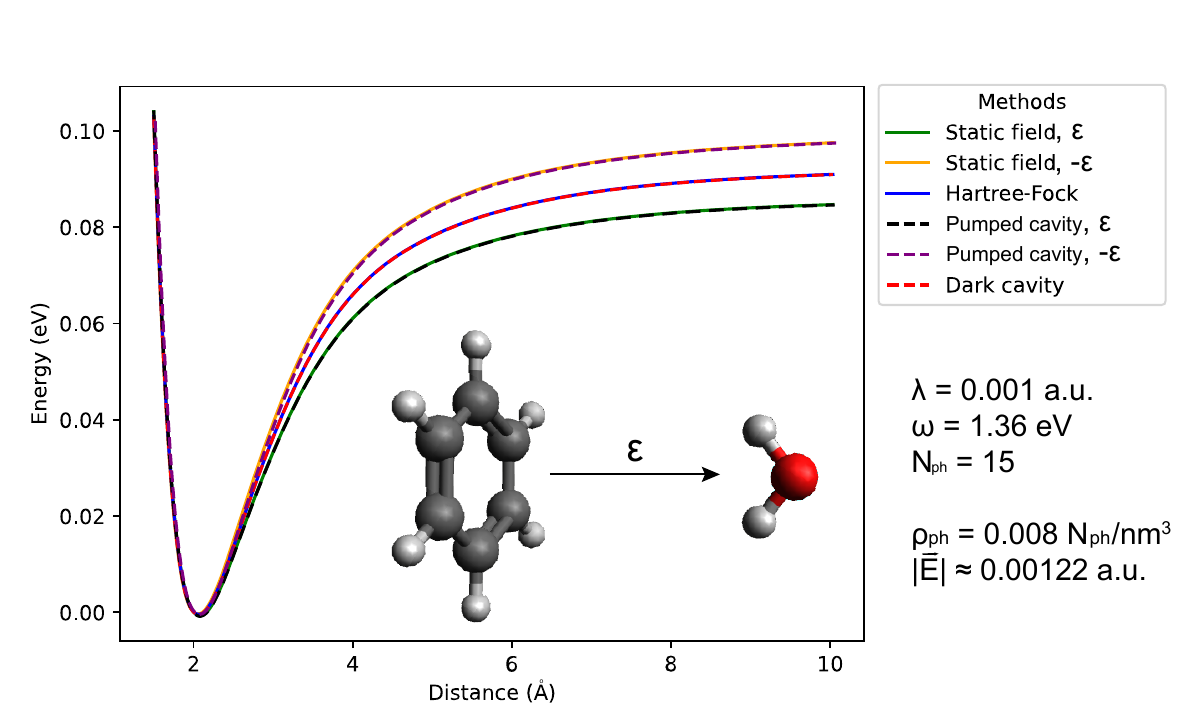}\hfill
\includegraphics[width=0.49\textwidth]{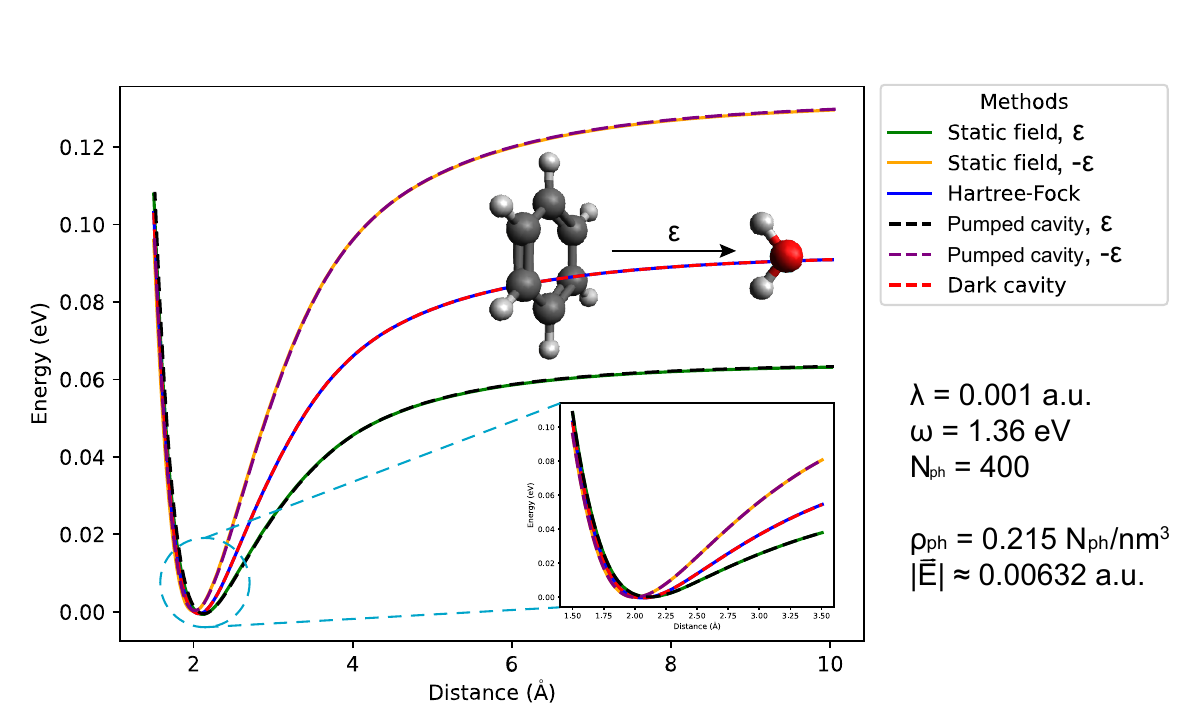}
\caption{Potential energy curves for the benzene-water complex as a function of the intermolecular distance, in a pumped cavity with $\lambda=0.001$~a.u.\ and $\omega=1.36$~eV. To the left $N_{\mathrm{ph}}=15$ and to the right $N_{\mathrm{ph}}=400$, with the field along $+\pmb{\epsilon}$, pointing from benzene to water, and along $-\pmb{\epsilon}$, pointing to the reverse.
The pumped cavity is compared with static fields of the same strength, with Hartree-Fock, and with the dark cavity.
Each curve is shifted so that its minimum is zero; the inset magnifies the region around the minimum.}
\label{fig:benzene_water}
\end{figure*}

The field also modifies intermolecular interactions.
Figure~\ref{fig:benzene_water} shows the potential energy curve of the benzene-water complex as a function of the intermolecular distance, with the field polarized along the axis connecting the two molecules, for $N_{\mathrm{ph}}=15$ and $400$ at $\lambda=0.001$~a.u.
Each curve is shifted so that its minimum is zero.
The dark cavity curve coincides with the Hartree-Fock one: at the mean field level and at this coupling strength, the vacuum field has no visible effect on the interaction.
Cavity-induced modifications of intermolecular interactions, including interactions that persist at large separation, arise from electron-photon correlation and require correlated methods such as QED coupled cluster~\cite{haugland2021intermolecular} or strong coupling M{\o}ller-Plesset perturbation theory~\cite{el2025strong}.
With $N_{\mathrm{ph}}=15$ the binding energy of about 0.09~eV changes by a few meV, while with $N_{\mathrm{ph}}=400$ it decreases to about 0.06~eV when the field points from benzene to water, along $+\pmb{\epsilon}$, and increases to about 0.13~eV when it points from water to benzene.
The asymmetry reflects the interaction between the dipole induced in benzene and the permanent dipole of water, where the hydrogen atoms point towards the ring~\cite{suzuki1992benzene}: along $+\pmb{\epsilon}$ the induced dipole is antiparallel to that of water and the interaction is repulsive, while along $-\pmb{\epsilon}$ the two dipoles are parallel and the interaction is attractive.
For both photon numbers and both orientations the pumped cavity and the static field of the same strength give the same curves.

\begin{figure*}[t]
\centering
\includegraphics[width=0.7\textwidth]{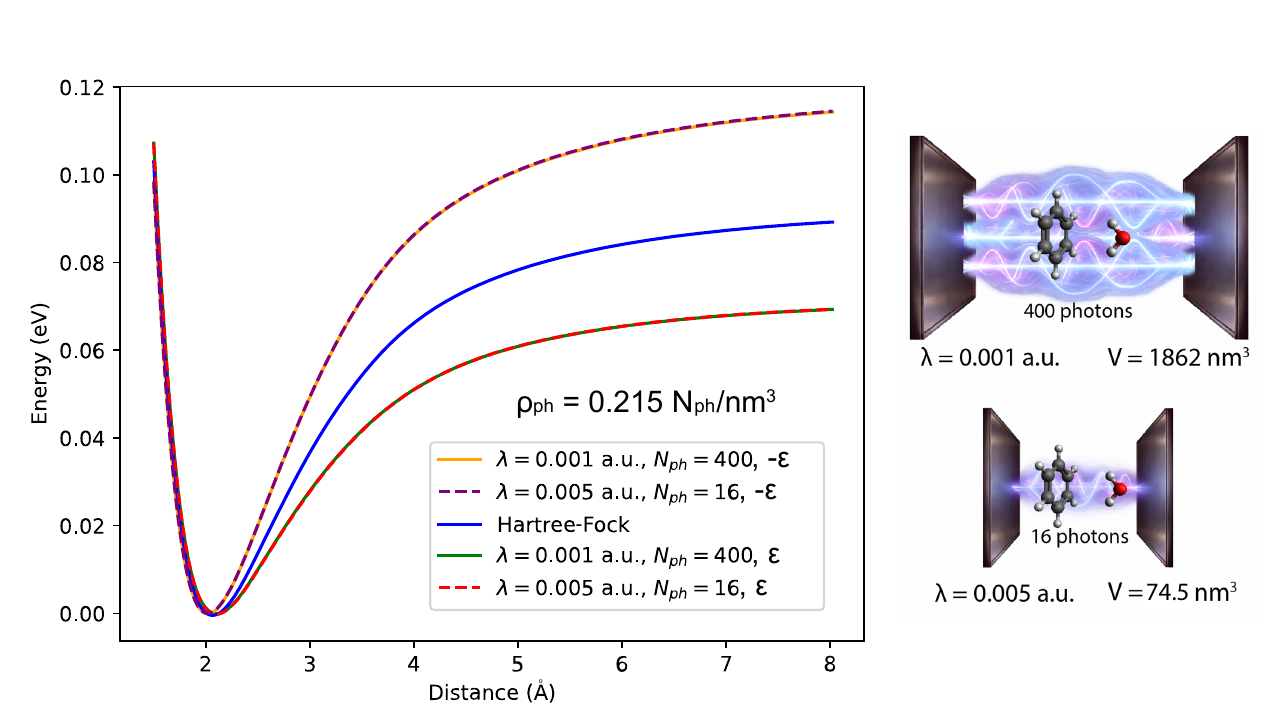}
\caption{Potential energy curves of the benzene-water complex in two cavities with the same photon density $\rho_{\mathrm{ph}}=0.215$~nm$^{-3}$: $\lambda=0.001$~a.u.\ with $N_{\mathrm{ph}}=400$ ($V=1862$~nm$^3$) and $\lambda=0.005$~a.u.\ with $N_{\mathrm{ph}}=16$ ($V=74.5$~nm$^3$), for the field along $+\pmb{\epsilon}$ and $-\pmb{\epsilon}$ and with $\omega=1.36$~eV, compared with Hartree-Fock.}
\label{fig:photon_density}
\end{figure*}

\subsection{Coupling strength and photon number}\label{sec:photon_density}
Since $\lambda^2=4\pi/V$, the field of Eq.~\ref{eq:field_classical} depends on the coupling strength and on the photon number only through the photon density $\rho_{\mathrm{ph}}=N_{\mathrm{ph}}/V$,
\begin{equation}\label{eq:photon_density}
E_0 \simeq \lambda\sqrt{2\omega N_{\mathrm{ph}}} = \sqrt{8\pi\omega\rho_{\mathrm{ph}}} ,
\end{equation}
so that the energy density of the classical field, $E_0^2/8\pi$, equals the energy density of the photons, $\omega\rho_{\mathrm{ph}}$.
Cavities that differ in volume but hold the same photon density should therefore have the same effect on the molecule.
Figure~\ref{fig:photon_density} confirms this for two cavities with $\rho_{\mathrm{ph}}=0.215$~nm$^{-3}$: one with $\lambda=0.001$~a.u.\ and $N_{\mathrm{ph}}=400$, and the other with $\lambda=0.005$~a.u.\ and $N_{\mathrm{ph}}=16$, where the volumes differ by a factor of 25.
The benzene-water curves coincide for both orientations of the field.
A large cavity with a weak vacuum coupling that holds many photons thus acts on the molecule as a small cavity holding few, and pumping reaches the same field effects without shrinking the mode volume.
The equivalence between cavities with the same photon density is not exact due to the polarization term $\lambda^2\langle\mathbf{d}\cdot\pmb{\epsilon}\rangle$ in Eq.~\ref{eq:field_classical}, which is independent on the photon number, and to the effects of the vacuum field already present in the dark cavity.
The coincidence of the curves in Figure~\ref{fig:photon_density} shows that these contributions are negligible at the coupling strengths considered here, and that the effect of the pumped cavity is determined by the classical part of the field.
Conversely, a difference between cavities with the same photon density would isolate the contribution of the quantized field, which is expected to grow with the coupling strength.
In the static bias interpretation, the photon density therefore sets the strength of an oriented external electric field, which is known to control molecular structure and reactivity~\cite{shaik2018structure,aragones2016electrostatic,hoffmann2022linear}.

\subsection{Torsional potential and infrared spectrum of hydrogen peroxide}\label{sec:h2o2}
\begin{figure}[t]
\centering
\includegraphics[width=\columnwidth]{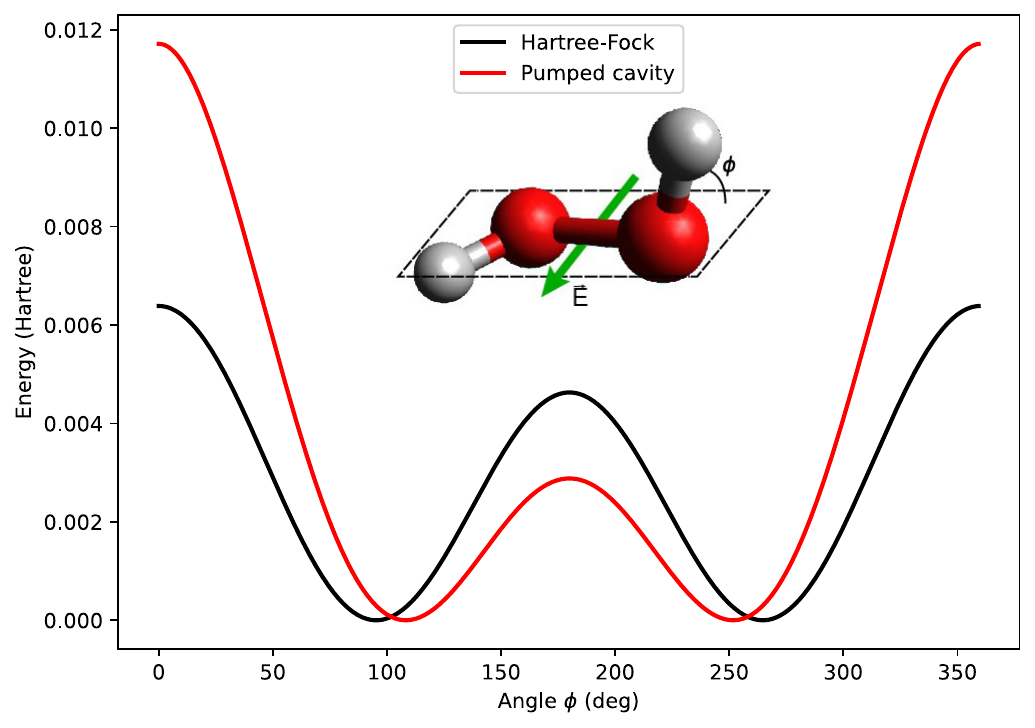}
\caption{Torsional potential of hydrogen peroxide as a function of the torsional angle $\phi$ ($\phi=0^\circ$ trans, $\phi=180^\circ$ cis), for Hartree-Fock (black) and in a pumped cavity with $\lambda=0.001$~a.u., $\omega=1.36$~eV and $N_{\mathrm{ph}}=400$ (red), averaged over the orientation of the field with Eq.~\ref{eq:boltzmann_potential}.
Both potentials are shifted so that their minima are zero.}
\label{fig:potential}
\end{figure}
\begin{table}[t]
\caption{Torsional barriers of hydrogen peroxide, measured from the minimum of the potential, and lowest torsional levels $E_n$, measured from the ground level $E_0$, for the Hartree-Fock and the pumped-cavity potentials of Figure~\ref{fig:potential}.}
\label{tab:torsion}
\centering
\begin{tabular}{lcc}
\toprule
 & Hartree-Fock & Pumped cavity \\
\midrule
Cis barrier (kcal/mol) & 2.91 & 1.81 \\
Trans barrier (kcal/mol) & 4.01 & 7.35 \\
Cis barrier (cm$^{-1}$) & 1016.2 & 632.6 \\
Trans barrier (cm$^{-1}$) & 1401.7 & 2571.0 \\
\midrule
$E_1-E_0$ (cm$^{-1}$) & 0.23 & 3.86 \\
$E_2-E_0$ (cm$^{-1}$) & 397.8 & 346.8 \\
$E_3-E_0$ (cm$^{-1}$) & 404.4 & 414.9 \\
$E_4-E_0$ (cm$^{-1}$) & 709.5 & 644.7 \\
$E_5-E_0$ (cm$^{-1}$) & 769.3 & 837.5 \\
$E_6-E_0$ (cm$^{-1}$) & 957.7 & 1067.3 \\
$E_7-E_0$ (cm$^{-1}$) & 1124.6 & 1304.1 \\
\bottomrule
\end{tabular}
\end{table}
Finally, we study how the pumped field modifies an intramolecular potential, the torsion of hydrogen peroxide around the O-O bond.
We compute the energy on a grid of torsional angles $\phi$ with a spacing of $10^\circ$ for $N_{\mathrm{ph}}=400$ and $\lambda=0.001$~a.u., corresponding to $\rho_{\mathrm{ph}}=0.215$~nm$^{-3}$ and to a electric field of $6.3\times10^{-3}$~a.u.
The angle $\phi$ is $180^\circ$ minus the H-O-O-H dihedral angle, so that $\phi=0^\circ$ corresponds to the trans and $\phi=180^\circ$ to the cis conformation. The performed scan is rigid: one hydrogen atom rotates around the O-O axis at fixed O-O and O-H bond lengths of 1.469~\AA{} and 0.967~\AA{} and a fixed O-O-H angle of $114.0^\circ$.
Since the molecule can orient freely with respect to the field, at each angle we average over the directions $\pmb{\epsilon}_i$ of the polarization on the unit sphere with Boltzmann weights at $T=298$~K,
\begin{equation}\label{eq:boltzmann_potential}
V(\phi) = \frac{\sum_i w_i\,E_i(\phi)\,e^{-\beta E_i(\phi)}}{\sum_i w_i\,e^{-\beta E_i(\phi)}} ,
\end{equation}
where $E_i(\phi)$ is the energy for the polarization $\pmb{\epsilon}_i$, $\beta=1/k_BT$ and $w_i=|\sin\theta_i|$, with $\theta_i$ the polar angle of $\pmb{\epsilon}_i$, is the solid-angle weight.
We interpolate the averaged energies with a periodic cubic spline.
In Figure~\ref{fig:potential} we compare the resulting potential with Hartree-Fock.
The pumped cavity lowers the cis barrier from 2.91 to 1.81~kcal/mol, raises the trans barrier from 4.01 to 7.35~kcal/mol (Table~\ref{tab:torsion}), and moves the minima towards the cis conformation.
This is the effect expected from the permanent dipole in the field: the dipole moment is largest in the cis conformation, which the orientational average stabilizes most, and vanishes by symmetry in the centrosymmetric trans conformation.
The torsional levels follow from the one-dimensional Hamiltonian
\begin{equation}\label{eq:torsion_hamiltonian}
H_{\mathrm{tor}} = -\frac{1}{2I}\frac{d^2}{d\phi^2} + V(\phi) ,
\end{equation}
where $I=m_{\mathrm{H}}r_\perp^2/2$ is the reduced moment of inertia of the two OH groups rotating around the O-O axis, $m_{\mathrm{H}}$ is the hydrogen mass, and $r_\perp=0.883$~\AA{} is the distance of the hydrogen atoms from the axis.
We solve the Schrödinger equation in the basis of plane waves $e^{im\phi}$, in which the kinetic energy is diagonal, on a Fourier grid of 512 points~\cite{marston1989fourier}.
In Hartree-Fock, tunneling through the cis barrier splits the ground state by 0.23~cm$^{-1}$ and the lower cis barrier in the cavity increases this splitting to 3.9~cm$^{-1}$.
The splitting of the first excited torsional doublet changes even more, from 7 to 68~cm$^{-1}$ (Table~\ref{tab:torsion}).
\begin{table}[b]
\caption{Frequencies (cm$^{-1}$) of the torsional transitions $m\to n$ of hydrogen peroxide that appear in the spectra of Figure~\ref{fig:spectra}, with the levels labeled as in Table~\ref{tab:torsion}.
Hot bands start from thermally populated excited levels.}
\label{tab:ir}
\centering
\begin{tabular}{llcc}
\toprule
 & $m\to n$ & Hartree-Fock & Pumped cavity \\
\midrule
Fundamental & $0\to2$ & 397.8 & 346.8 \\
 & $1\to2$ & 397.6 & 342.9 \\
 & $0\to3$ & 404.4 & 414.9 \\
 & $1\to3$ & 404.2 & 411.0 \\
\midrule
Hot band & $2\to3$ & 6.6 & 68.1 \\
 & $2\to4$ & 311.7 & 298.0 \\
 & $3\to4$ & 305.1 & 229.8 \\
 & $3\to5$ & 364.9 & 422.7 \\
 & $4\to5$ & 59.8 & 192.8 \\
 & $4\to6$ & 248.1 & 422.6 \\
 & $5\to7$ & 355.3 & 466.6 \\
\midrule
Overtone & $0\to4$ & 709.5 & 644.7 \\
\bottomrule
\end{tabular}
\end{table}
\begin{figure*}[t]
\centering
\includegraphics[width=0.49\textwidth]{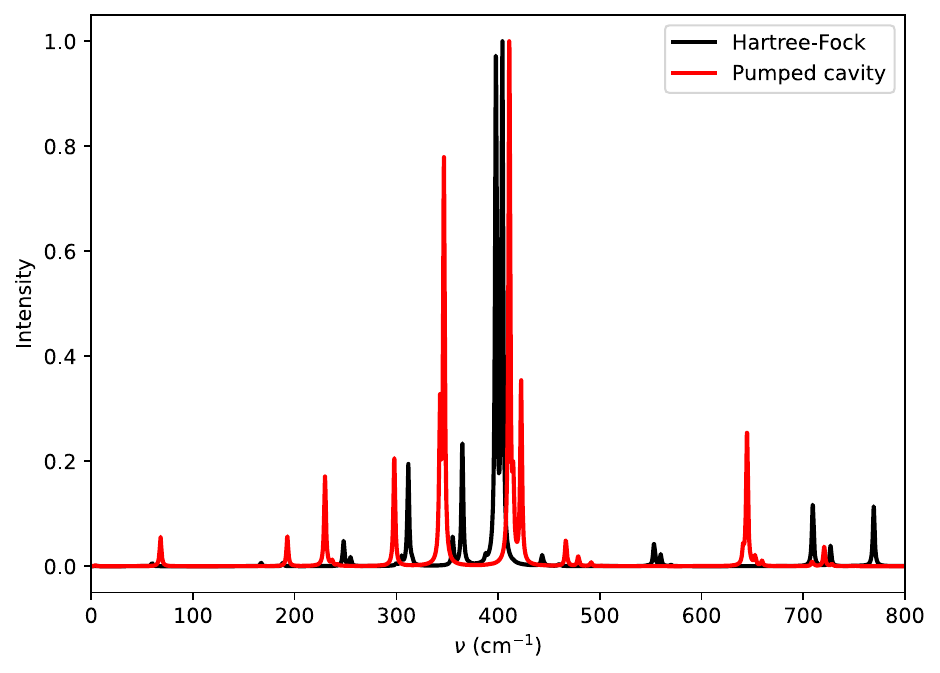}\hfill
\includegraphics[width=0.49\textwidth]{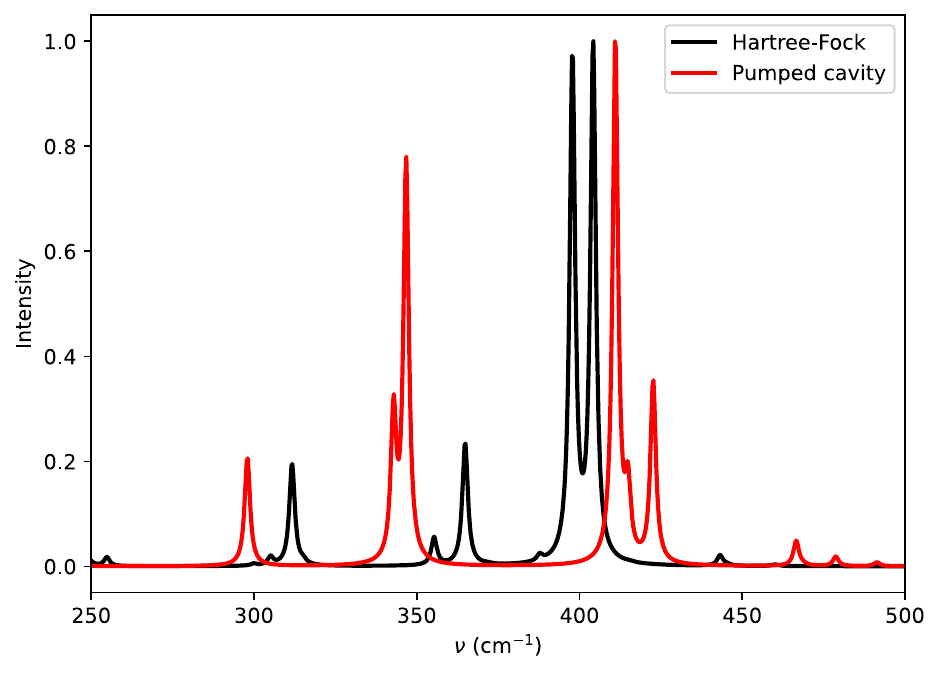}
\caption{Torsional absorption spectra of hydrogen peroxide at 298~K from the Hartree-Fock (black) and the pumped-cavity (red) potentials of Figure~\ref{fig:potential}, computed with Eq.~\ref{eq:spectrum}.
Left: 0--800~cm$^{-1}$.
Right: magnification of the 250--500~cm$^{-1}$ region.
Each spectrum is normalized to its most intense peak.}
\label{fig:spectra}
\end{figure*}
The torsional absorption spectrum follows from the oscillator strengths
\begin{equation}\label{eq:oscillator_strength}
f_{mn} = \frac{2}{3}\,(P_m-P_n)\,\omega_{mn}\,|\pmb{\mu}_{mn}|^2 ,
\end{equation}
where $\omega_{mn}$ is the transition frequency, $P_m$ is the Boltzmann population of level $m$ at 298~K, the difference $P_m-P_n$ accounts for stimulated emission, and $\pmb{\mu}_{mn}=\langle\psi_m|\pmb{\mu}(\phi)|\psi_n\rangle$ is the transition dipole.
Each transition is broadened by a Lorentzian of half width $\gamma=1$~cm$^{-1}$,
\begin{equation}\label{eq:spectrum}
I(\omega) = \sum_{m<n} f_{mn}\,\frac{\gamma^2}{(\omega-\omega_{mn})^2+\gamma^2} ,
\end{equation}
where the sum includes the lowest 15 torsional levels.
The dipole function $\pmb{\mu}(\phi)=\sum_AZ_A\mathbf{R}_A(\phi)$ is evaluated from the nuclear charges $Z_A$ and positions $\mathbf{R}_A$ and is the same for both potentials.
The spectra therefore report the line positions and the Boltzmann-weighted relative intensities and their differences originate entirely from the modified torsional potential.
Figure~\ref{fig:spectra} compares the two spectra, each normalized to its most intense line.
Table~\ref{tab:ir} lists the frequencies of the transitions that appear in the spectra.
In Hartree-Fock, the most intense lines at 398 and 404~cm$^{-1}$ connect the ground doublet with the first excited one, with hot bands at 311 and 365~cm$^{-1}$.
In the cavity the larger splitting of the excited doublet separates these transitions into two groups, near 343--347 and 411--415~cm$^{-1}$, the hot bands move to 298 and 423~cm$^{-1}$, where the $3\to5$ and $4\to6$ transitions overlap, and the transition within the excited doublet appears at 68~cm$^{-1}$.
The torsional spectrum thus provides a far-infrared fingerprint of the potential reshaped by the pumped cavity.
Signatures of this kind are accessible to the cavity-enhanced and pump-probe spectroscopies used to interrogate molecules in cavities~\cite{baradaran2026long,chen2025ultrafast}, where the molecular response must be separated from the optical effects of the cavity.

\section{Conclusions}\label{sec:conclusions}
We have introduced a mean field description of a molecule in a pumped optical cavity, in which a prescribed number of photons is imposed on the SC-QED-HF wave function through a Lagrange multiplier.
Since every eigenstate of the Pauli-Fierz Hamiltonian carries no transverse electric field, a cavity holding real photons and an electric field cannot be described as an excited state of the light-matter Hamiltonian. However the constrained reference provides the non-equilibrium state that the pump maintains.
The multiplier replaces the photon frequency by the effective free-field frequency $\omega+\mu$ and the electric field of the constrained reference follows in closed form from Eq.~\ref{eq:field_mu}, that is, it vanishes without pumping and approaches the field $\lambda\sqrt{2\omega N_{\mathrm{ph}}}$ of a coherent state when the variance of the displacement is negligible.
For benzene and the benzene-water complex, the pumped reference polarizes the molecule and modifies the intermolecular interaction as a static classical field of the same strength.
Since the field depends only on the photon density, cavities of different volume holding the same photon density give the same results, and a large cavity with many photons acts on a molecule as a small cavity with few.
For hydrogen peroxide, the field lowers the cis barrier and raises the trans barrier of the torsional potential, which increases the tunneling splittings and shifts the far-infrared torsional lines by tens of wavenumbers.
The classical driven field limit thus emerges from a quantized description, and the photon number provides a control parameter for cavity-modified chemistry that does not require shrinking the cavity mode quantization volume.
The enhancement of the light-matter interaction by pumping is complementary to the enhancement obtained by reducing the quantization volume, and it makes strong electric fields accessible in cavities of larger volume, such as the high mode number cavities used for electronic strong coupling in solution~\cite{kushida2025fluidic} or the high finesse cavities of cavity-enhanced molecular spectroscopy~\cite{baradaran2026long}.
External driving tunes photon-mediated interactions in the same way in cavity QED platforms~\cite{hosseinabadi2026kinetically}.

Several aspects of the present treatment can be improved.
First, the constrained reference describes a snapshot of the system at the maximum of the oscillating field.
This is adequate for the electrons when the cavity frequency lies below the electronic excitation energies, but the nuclei respond according to the Born-Oppenheimer approximation.
In an optically pumped cavity the interaction between the permanent dipole and the electric field averages to zero over a cycle, and a cycle-averaged or Floquet treatment~\cite{shirley1965solution,sentef2020quantum} would be required to describe the modification of intermolecular interactions and torsional barriers.
Second, constraining the photon number, which is quadratic in the photon operators, drives the effective free-field frequency of the mode towards zero and flattens the photonic curvature in Eq.~\ref{eq:hessian_block}.
A constraint on the field coordinate $\langle b+b^\dagger\rangle$, which is linear in the photon operators, would impose the field without modifying the frequency of the mode, and it is equivalent to a static electric field applied to a molecule.
Such a combination of a classical and a quantized fields, already used to compute static polarizabilities in cavities~\cite{deprince2025static}, is a natural extension of the present work.
Third, the calculations are performed at the mean field level.
Electron-electron correlation and the electron-photon correlation beyond the orbital-specific displacement are not included, while they are responsible for the cavity-induced modifications of intermolecular interactions that are not visible in the dark cavity~\cite{haugland2021intermolecular}.
Strong coupling M{\o}ller-Plesset perturbation theory~\cite{el2025strong} and QED coupled cluster~\cite{haugland2020coupled} built on the pumped reference would include these effects.
Finally, the torsional spectrum of hydrogen peroxide was obtained from a rigid one-dimensional scan, in which the torsion is treated as an isolated coordinate and the transition dipoles are evaluated from the nuclear charges only.
The torsion is coupled to the other vibrational modes of the molecule, such as the O-O-H bends, and a multidimensional treatment with relaxed geometries and the full dipole surface would be needed for a quantitative comparison with experiment~\cite{hunt1965internal}.

The constrained reference also opens further directions.
Response theory~\cite{castagnola2024polaritonic,castagnola2025strong,yuwono2026dynamic} built on the Lagrangian of Eq.~\ref{eq:lagrangian} would give access to the polaritonic excitations of a pumped cavity and to the photon number dependence of the Rabi splitting discussed in Section~\ref{sec:introduction}.
This requires care, because the constraint changes the effective free-field frequency of the mode from $\omega$ to $\omega+\mu$.
The photon statistics of the constrained reference and its extension to squeezed states, which is required to describe nonclassical light, are discussed in Appendices~\ref{app:statistics} and~\ref{app:squeezing}.
Finally, extending the approach to several cavity modes would allow the study of multimode effects~\cite{hoffmann2020effect}, which are neglected in the single-mode description used here.


\section{Funding}
YEM, RRR and HK acknowledge funding from the European Research Council (ERC) under the European Union’s Horizon 2020 Research and Innovation Programme (grant agreement No. 101020016).

\section{Acknowledgement}
We acknowledge insightful discussions with Lorenzo D. Rossi.

\section{Author contributions}
YEM, RRR and HK conceived the project.
YEM carried out the theoretical modeling, the implementation of the method and performed the numerical simulations.
HK supervised the project.
All authors discussed the results and revised the manuscript.

\section{Data availability}
The e$^\mathcal{T}$ code and all the results can be made available upon reasonable request to the authors.

\section{Conflict of interest}
Authors state no conflict of interest.

\appendix

\section{Photon statistics}\label{app:statistics}
Beyond the field experienced by the molecule, the constrained reference also determines the statistics of the cavity photons~\cite{brorsen2026intracavity}, which characterize whether the light it describes is classical.
In the dipole basis, the SC-QED-HF reference is a superposition of determinants $|I\rangle$,
\begin{equation}
    \sum_Ic_I|I\rangle\otimes|\alpha_I\rangle ,
\end{equation}
each dressed by a coherent state with displacement
\begin{equation}
    \alpha_I= \frac{\lambda}{\sqrt{2\omega}} x_I ,
\end{equation}
where $x_I$ is the eigenvalue of $X=\sum_p\eta_p\tilde{E}_{pp}$ on $|I\rangle$.
Tracing out the electrons leaves a statistical mixture of coherent states with weights $|c_I|^2$, and the photon number $n=b^\dagger b$ has the variance
\begin{equation}\label{eq:photon_variance}
\sigma^2 = \langle n^2\rangle-\langle n\rangle^2 = \langle n\rangle + \mathrm{Var}_I\big(\alpha_I^2\big) ,
\end{equation}
where the first term is the shot noise of each coherent state and the second is the spread of the displacements over the determinants, and $\mathrm{Var}(n)$ denotes this variance.
The following optical quantities
\begin{equation}\label{eq:optical_quantities}
\begin{alignedat}{2}
&F = \mathrm{Var}(n)/\langle n\rangle , &\quad& \text{Fano factor} \\
&Q = F-1 , && \text{Mandel parameter} \\
&g^{(2)}(0) = 1+Q/\langle n\rangle , && \text{second-order coherence}
\end{alignedat}
\end{equation}
therefore satisfy $F\geq1$, $Q\geq0$ and $g^{(2)}(0)\geq1$ for any choice of the coherent state parameters~\cite{mandel1979sub,walls2008quantum}.
The light described by the reference is thus classical by construction, since a mixture of coherent states has a non-negative Glauber-Sudarshan $P$ function~\cite{sudarshan1963equivalence,glauber1963coherent}, consistent with the classical driven field limit found in Section~\ref{sec:results}.
The spread of the displacements vanishes when all $\eta_p$ are equal, as in QED Hartree-Fock, and it measures the electron-photon correlation carried by the orbital-specific parameters.
These quantities are currently implemented and follow from the cumulants of $X$ over the reference determinant, the first of which is the mean displacement of Eq.~\ref{X_avg}.
Analysis of \ref{eq:optical_quantities}, in particular for the unpumped ground state, will be the subject of future work.

\section{Squeezing}\label{app:squeezing}
Nonclassical light requires going beyond coherent states.
A squeezing transformation of the form
\begin{equation}
    U_{\mathrm{sq}}=\exp[\gamma(b^{\dagger2}-b^2)],
\end{equation}
with real $\gamma$, transforms the photon operators as
\begin{equation}\label{eq:squeezed_b}
U_{\mathrm{sq}}^\dagger b\,U_{\mathrm{sq}} = b\cosh(2\gamma) + b^\dagger\sinh(2\gamma) ,
\end{equation}
such that the field quadrature that couples to the molecule is rescaled as
\begin{equation}\label{eq:squeezing}
U_{\mathrm{sq}}^\dagger(b+b^\dagger)U_{\mathrm{sq}} = e^{2\gamma}(b+b^\dagger) .
\end{equation}
The variance of the photonic coordinate $x$ and its conjugated momentum $p$, given by
\begin{equation}
\begin{split}
    &x=(b+b^\dagger)/\sqrt{2} \\
    &p=i(b^\dagger-b)/\sqrt{2} ,
\end{split}
\end{equation}
change from the vacuum value $1/2$ to $e^{4\gamma}/2$ and $e^{-4\gamma}/2$ respectively.
Applying the transformation to Eq.~\ref{eq:pauli_fierz} gives the squeezed Pauli-Fierz Hamiltonian
\begin{equation}\label{eq:squeezed_pf}
\begin{split}
H_{\mathrm{sq}} &= U_{\mathrm{sq}}^\dagger H\,U_{\mathrm{sq}} \\
&= H_e - e^{2\gamma}\lambda\sqrt{\frac{\omega}{2}}(\mathbf{d}\cdot\pmb{\epsilon})(b+b^\dagger) \\
&+ \frac{\lambda^2}{2}(\mathbf{d}\cdot\pmb{\epsilon})^2 + \omega\cosh(4\gamma)\,b^\dagger b \\
&+ \frac{\omega}{2}\sinh(4\gamma)\big(b^2+b^{\dagger2}\big) + \omega\sinh^2(2\gamma) .
\end{split}
\end{equation}
The bilinear coupling is multiplied by $e^{2\gamma}$, while the dipole self energy, which acts on the electrons only, is unchanged.
For $\gamma<0$ the coordinate $x$, which couples to the molecule, is squeezed below the vacuum level, which is a genuinely nonclassical state, but the light-matter coupling is reduced.
For $\gamma>0$ the coupling is enhanced, as in previous works that amplify light-matter interactions through antisqueezing~\cite{leroux2018enhancing,qin2018exponentially}, while $p$ is squeezed below the vacuum level, such that the light is nonclassical in both cases.
The enhancement increase the complexity in the photonic part of Eq.~\ref{eq:squeezed_pf}, whose vacuum expectation value $\omega\sinh^2(2\gamma)$ is the energy of the $\sinh^2(2\gamma)$ photons of the squeezed vacuum, while the terms in $b^2$ and $b^{\dagger2}$ create and annihilate photon pairs.
The squeezing only acts on the photons and therefore commutes with the orbital rotations, but not with the orbital-specific displacement of Eq.~\ref{eq:U_SC}.
A squeezed SC-QED-HF wave function can be defined as
\begin{equation}\label{eq:sq_wf}
|\psi_{\mathrm{sq}}\rangle = U_{\mathrm{sq}}\,U_{\mathrm{SC}}\,e^{\kappa}|\mathrm{HF}\rangle\otimes|0\rangle ,
\end{equation}
with an energy given by the expectation value of the squeezed Hamiltonian of Eq.~\ref{eq:squeezed_pf} over the SC-QED-HF wave function of Eq.~\ref{eq:sc_wf}.
Since $U_{\mathrm{SC}}$ displaces the mode by an amount that depends on the orbital occupations, moving the squeezing through only rescales the displacement,
\begin{equation}\label{eq:sq_order}
U_{\mathrm{sq}}\,U_{\mathrm{SC}}(\pmb{\eta}) = U_{\mathrm{SC}}(e^{2\gamma}\pmb{\eta})\,U_{\mathrm{sq}} ,
\end{equation}
where $U_{\mathrm{SC}}(\pmb{\eta})$ is Eq.~\ref{eq:U_SC} with parameters $\pmb{\eta}$.
Although the two transformations do not commute, the two orderings describe the same set of states and they give the same optimized energy.
In terms of the rescaled parameters
\begin{equation}
    \eta'_p=e^{2\gamma}\eta_p,
\end{equation}
the energy is given by Eq.~\ref{eq:E_SC} with $\eta'$ in place of $\eta$ and with the Gaussian factors
\begin{equation}\label{eq:sq_gaussian}
\begin{split}
&Q_{pq} = \exp\Big(-\frac{\lambda^2}{4\omega e^{4\gamma}}\Delta_{pq}'^{\,2}\Big) \\
&Q_{pqrs} = \exp\Big(-\frac{\lambda^2}{4\omega e^{4\gamma}}\Delta_{pqrs}'^{\,2}\Big) ,
\end{split}
\end{equation}
plus the energy $\omega\sinh^2(2\gamma)$ of the squeezed vacuum. The photon number, on the other hand, reads
\begin{equation}\label{eq:sq_photon_number}
\langle b^\dagger b\rangle_{\mathrm{sq}} = \frac{\lambda^2}{2\omega}\sum_{pq}\eta'_pS_{pq}\eta'_q + \sinh^2(2\gamma) .
\end{equation}
The electric field is given by Eq.~\ref{eq:field} with $\eta'$ in place of $\eta$, so the photons of the squeezed vacuum carry no field.
Squeezing thus enters the energy through the Gaussian factors, on which it acts as the rescaling $\omega\to\omega e^{4\gamma}$ of the cavity frequency.
The sign of $\gamma$ has then an effect on the damping of the one and two-electron integrals through the Gaussian factors.
For a common $\eta$, as in the QED Hartree-Fock, all Gaussian factors equal one and $\gamma$ enters only through the energy and the photons of the squeezed vacuum.
With orbital-specific parameters, the Gaussian factors depend on $\gamma$ already in first order, while the energy and the photon number of the squeezed vacuum depend in second order in $\gamma$.
The Fano factor, the Mandel parameter and $g^{(2)}(0)$ do not detect this kind of nonclassicality, since the squeezed vacuum has $F=2(1+\langle n\rangle)$ and the coordinate/momentum variance are the true observables of such effects~\cite{walls2008quantum}.
Imposing a condition such that the squeezing/antisqueezing is kept throughout the optimization of the variational parameters will be the subject of a dedicated investigation.

\bibliography{main}

\end{document}